\documentclass[aps,prb,twocolumn,showpacs,amsfonts,amsmath,amssymb]{revtex4}
\usepackage{eucal}
\usepackage{cmmib57}
\usepackage{txfonts}  
\usepackage{textcomp} 
\usepackage{graphics}
\usepackage{epsfig}

\begin{document}
   \title{
Persistent currents in ballistic normal-metal rings
         }
\author{
Michael Moskalets
}
\affiliation{
Department of Metal and Semiconductor Physics, NTU "Kharkiv Polytechnic Institute", 61002 Kharkiv, Ukraine
}
\date\today
   \begin{abstract}
Recent experiments
renewed interest in persistent currents in mesoscopic normal-metal rings.
We show that in ballistic rings in high magnetic fields the Zeeman splitting leads to periodic current quenching with period much larger than the period of the persistent current.
Simple arguments show that this effect might be relevant for diffusive rings as well.
Another aim of this paper is to discuss fluctuations of the persistent current due to thermal excitation of high energy levels.
Being observed such fluctuations would witness a coherent state of an electron system at high temperatures when the persistent current is exponentially suppressed.
\end{abstract}
\maketitle

\section{Introduction}
\label{sec1}
Recent experiments \cite{Bleszynski-Jayich09, Bluhm09} renewed interest in persistent currents in mesoscopic normal-metal rings.
The existence of non-decaying (persistent) currents in rings pierced by a magnetic flux was predicted for ballistic \cite{Kulik70} as well as for more realistic rings with elastic scattering \cite{BIL83}.
Experiments using ballistic rings \cite{MCB93, RSMHBE01}  agree well with theoretical predictions \cite{Kulik70, CGRS88}. In contrast diffusive rings provide a longstanding challenge due to an apparent disagreement between experiment and theory.
The experiment finds \cite{LDDB90, CWBKGK91, JMKW01} a current a few order of magnitude larger than the theory \cite{RO93} based on a model of non-interacting diffusive electrons.
More refined theory, for example accounting for electron-electron interactions \cite{AE90, ES95}, could not remove this disagreement.
However the new experiments \cite{Bleszynski-Jayich09, Bluhm09} made with the help of more sensitive techniques showed an amassing agreement with predictions of the non-interacting theory \cite{RO93}
including Zeeman splitting and spin-orbit scattering \cite{GGOOSBH09}.
In the recent  experiment of Bleszynski-Jayich {\it et al.}  \cite{Bleszynski-Jayich09} this agreement is possible a consequence of the high magnetic field that penetrates the ring and that suppresses weak localization and related interaction effects. The good agreement opens the door to use persistent currents as a tool to provide direct information on the quantum state of closed systems of electrons. The excellent agreement reinforces us that the theory of non-interacting electrons remains a powerful theory in mesoscopic physics.

In the case of diffusive rings the quantity which is usually discussed is a typical persistent current, a square root of a mean square current.
In theory averaging is performed over disorder potentials.
In experiment averaging is performed over a relevant interval of magnetic fields.
This averaging is over static fluctuations similar in nature to the universal conductance fluctuations \cite{Altshuler85, LS85, WW86}.
However there are also intrinsic fluctuations (time-dependent noise) of persistent currents at zero temperature \cite{BS96, CB98, CPB00, BJ05, SZ10} as well as at finite temperatures \cite{Moskalets01}$^{,}$ \cite{SZ10}.

One aim of this paper is to present a short survey of the theoretical results on intrinsic persistent current fluctuations.

The other aim of this paper is related to an interesting
effect found in the experiment by Bleszynski-Jayich {\it et al.} \cite{Bleszynski-Jayich09}: The persistent current is quenched periodically with increasing magnetic field, see Fig.~1E in Ref.~\onlinecite{Bleszynski-Jayich09}.
The corresponding period, $\Delta B\sim 0.3$\,T, is close to a magnetic field increment $\Delta B_{Z}$ necessary to increase the Zeeman splitting of the order of the Thouless energy $E_{Th}$.
For a diffusive ring of length $L$ we have $E_{Th} = \pi^2\hbar D/L^2$ with $D$ the diffusion constant. \cite{RO93}
For typical rings' parameters of Ref.~\onlinecite{Bleszynski-Jayich09}, $L \sim 2000$~nm and
$D \sim 270$~cm$^2$/s the Thouless energy corresponds to a temperature $T_{Th} = E_{Th}/k_{B} \sim 0.5$~K,
with $k_{B}$ the Boltzmann constant.
With this we find $\Delta B_{Z} = E_{Th}/(g\mu_{B}) \sim 0.37$~T where we have used the gyromagnetic ratio $g = 2$ and the Bohr magneton $\mu_{B} = e\hbar/(2m_{e})$ calculated with the free electron mass $m_{e}$.
Since $\Delta B \approx \Delta B_{Z}$, one can conjecture that the periodic increment of Zeeman splitting by the Thouless energy results in periodic persistent current quenching.
It can be understood as follows:
In the diffusive ring of length $L$ with $N$ transverse channels and an electron mean free path $l$  one can arrange levels into groups containing $N_{eff} = N l/L$ correlated levels for each spin direction.
Each such a group spans an energy window of order the Thouless energy $E_{Th}$.
With increase of magnetic field the relative position (in energy) of spin-up and spin-down level groups is varied with period $\Delta B_{Z}$ due to Zeeman energy.
This results in a corresponding periodicity of the persistent current magnitude.
How this mechanism works precisely we show for a single channel ballistic ring when the Thouless energy equals the level spacing $\Delta_{F}$ near the Fermi energy and the effective number of channels is $N_{eff} = 1$.

The paper is organized as follows: In Sec.~\ref{sec2} we explore the effect of a high magnetic field onto the persistent current in ballistic rings. We show that the fluctuations in spin subsystem, taking place at some particular magnetic fields, dramatically reduces the persistent current magnitude at finite temperatures.
Then in Sec.~\ref{sec3} we discuss fluctuations of the persistent current in a single ring due to coupling to a thermal bath. We conclude in Sec.~\ref{sec4}.

\section{Effect of Zeeman splitting in ballistic rings}
\label{sec2}

The magnetic field $B$ has a twofold effect.
First, it produces an Aharonov-Bohm (AB) magnetic flux \cite{AB59} through the ring.
This results in a periodicity of the free energy with magnetic field with period $\Delta B_{AB} = \Phi_{0}/S$, where $\Phi_{0} = h/e$ is the magnetic flux quantum and $S$ is the area enclosed by the ring.\cite{BY61}
This periodicity is due to intersections (direct or avoided-crossing) of an electron spectrum sub-bands corresponding to different orbital moments. We assume that a reservoir keeps the system in the energetically most favorable state.

Second, the Zeeman splitting, which increases with $B$, leads to intersections of energy levels of electrons with spin up ($\uparrow$) and spin down ($\downarrow$).

If the spin-flip processes are present in the ring,
then the number of electrons with spin directed along (opposite to) the field will change as the field is varied.
This also results in oscillations of the thermodynamic quantities as a function of the magnetic fields but with a period of \cite{Moskalets99}

\begin{equation}
\label{eq00}
\Delta B_{Z} = \frac{ \Delta_{F} }{g \mu_{B} } \,.
\end{equation}

\noindent
For a ring with many electrons this period is much larger than the period of AB oscillations.

\subsection{Model}

\begin{figure}[b]
\centerline{\psfig{file=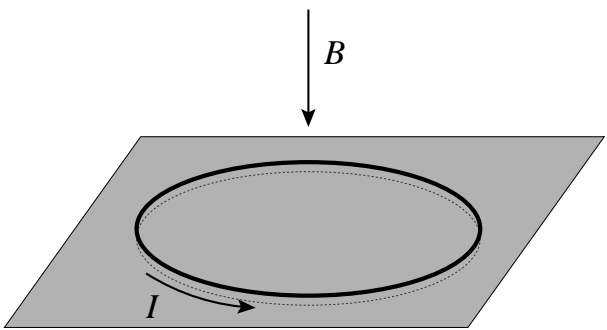,width=8.0cm}}
\caption{One-dimensional ring pierced by the magnetic field $B$ with persistent current $I$. The hatched plane represents for a reservoir of spin excitations which is uniformly coupled to the ring.}
\label{fig1}
\end{figure}

We consider a one-dimensional ($1D$) ballistic ring with non-interacting electrons in a perpendicular magnetic field, Fig.~\ref{fig1}.
Our calculations do not apply to diffusive rings and thus do not  describe quantitatively the results of experiments by Bleszynski-Jayich {\it et al.} \cite{Bleszynski-Jayich09}.

In real diffusive rings there is an additional effect which arises from the magnetic field penetrating the ring and changing the ring orbital wave functions. This effect can be expected to be periodic with the period required by increasing the flux into the sample by one flux quantum. Typically to see the Aharonov-Bohm flux period in a ring the ratio of the rings hole to the surface of the ring must be large. The interplay of these two orbital periodicities might also lead to quenching of the persistent current.

To model the presence of spin-flip processes we adopt the fictitious reservoir model introduced in Ref.~\onlinecite{Buttiker86}.
In our case it is a fictitious reservoir of spin excitations.
We assume that the ring can exchange electrons with a
reservoir having constant chemical potential $\mu_{0}$ independent of both the spin and magnetic field.
The chemical potential is positioned in the middle between electron levels of the ring in zero magnetic field.
Then with increasing $B$ energy levels for spin up electrons will decrease while the ones for spin down electrons will increase.
When some unoccupied level sinks below $\mu_{0}$ one electron with spin up enters the ring.
Similarly, when some occupied level rises above $\mu_{0}$ one electron with spin down escapes to the reservoir.
At chosen $\mu_{0}$ both crossings take place at the same magnetic field.
Hence the number of electrons in the ring remains fixed, that is the case for isolated rings used in experiment, while one spin is flipped.
To forbid charge fluctuations in the ring at finite temperature we additionally assume that the change of the particle number in the ring costs a large Coulomb energy $E_{c} \to \infty$. \cite{Moskalets99spin}

   \subsection{Main equations}

To describe a system of $N_{e} \gg 1$ non-interacting electrons in $1D$ ballistic ring we use the Luttinger liquid model \cite{Haldane81} with the Lagrangian in a bosonic form \cite{KF92},

\begin{equation}
\label{eq01}
L_{LL} = \dfrac{\hbar}{4} \sum\limits_{\chi = \rho, \sigma } \dfrac{1}{v_{F} }\, \left(\dfrac{ \partial\theta_{\chi} }{\partial t }  \right)^2 - v_{F}\, \left(\dfrac{ \partial\theta_{\chi } }{\partial x }  \right)^2 \,,
\end{equation}

\noindent
where $v_{F}$ is the Fermi velocity.
The fields $\theta_{\rho}$ and $\theta_{\sigma}$ describe charge and spin excitations with density $\rho_{\chi }(t,x) = \sqrt{1/\pi}\, \partial\theta_{\chi}/\partial x$ and flow $j_{\chi}(t,x) = \sqrt{1/\pi}\, \partial \theta_{\chi }/\partial t$, respectively.
The total number of electrons in the ring is $N_{e} \equiv N_{\uparrow} + N_{\downarrow} = \int_{0}^{L} \rho_{\rho}\, dx + N_{0}$ with $N_{0} = N_{0\uparrow} + N_{0\downarrow}$ the number of electrons in the ground state, i.e., at zero temperature and at $B = 0$.
Correspondingly, the number of spin excitations is $N_{\sigma} \equiv N_{\uparrow} - N_{\downarrow}  = \int_{0}^{L} \rho_{\sigma}\,dx + N_{0\sigma}$, where $N_{0\sigma} = N_{0\uparrow} - N_{0\downarrow}$ describes the spin polarization of the ground state.
In our model we have $N_{0\sigma} = 0$.

The presence of a magnetic field $B$ results in the Aharonov-Bohm phase and in the Zeeman energy.
The effect of the AB phase due to a magnetic flux $\Phi = B S$ through the ring is described by the Lagrangian \cite{Loss92,FK93}

\begin{equation}
\label{eq02}
L_{AB} = \dfrac{h }{L } \left\{\left[\dfrac{k_{j\rho} }{4 } + \dfrac{\Phi }{\Phi_{0} } \right]\, j_{\rho }  + \dfrac{k_{j\sigma} }{4 }\, j_{\sigma}  \right\},
\end{equation}

\noindent
where $L = \sqrt{4\pi S}$ is the circumference of the ring.
The topological numbers are $k_{j\rho} = k_{j\uparrow} + k_{j\downarrow}$ and $k_{j\sigma} = k_{j\uparrow} - k_{j\downarrow}$.
The spin-resolved topological numbers depend on the parity of the number of electrons in the ring: $k_{j\uparrow/\downarrow} = 0$ $(1)$ if $N_{\uparrow/\downarrow}$ is odd (even).
The Lagrangian, $L_{Z}$ which takes into account the Zeeman energy reads,
\begin{equation}
\label{eq03}
L_{Z} = g\mu_{B} B\, \dfrac{ N_{\sigma}}{L } \,.
\end{equation}

The particle exchange with a fictitious reservoir is described as follows, \cite{Moskalets00}

\begin{equation}
\label{eq03_01}
L_{ex} = \mu \rho_{\rho} \,-\, \dfrac{E_{c} }{L } \left( N_{e} - N_{0} \right)^2  \,.
\end{equation}

\noindent
At $E_{c} \to \infty$ the number of electrons (i.e., the charge) in the ring is frozen while the spin exchange with the reservoir is allowed.

With these Lagrangians we calculate the Euclidean action,

\begin{equation}
S_E=-\int\limits^{L }_{0} dx \int\limits^{\beta}_0 d\tau \left\{ L_{LL} + L_{AB} + L_{Z} + L_{ex} \right\} ,
\label{eq04}
\end{equation}

\noindent
where $\tau=it$ is an imaginary time and $\beta = \hbar/(k_{B} T)$ with $T$ temperature.

Then we calculate the partition function $Z$ as the path integral over the fields $\theta_{\rho}$ and $\theta_{\sigma}$,

\begin{equation}
Z = \int D \theta_{\rho}\, D\theta_{\sigma\,} e^{-\,\frac{S_{E} }{\hbar} }\,.
\label{eq05}
\end{equation}

\noindent
The partition function defines the thermodynamic potential $\Omega = -\, k_{B} T \ln{Z}$, which in turn defines the persistent current, \cite{BY61}

\begin{equation}
\label{pc}
I(\Phi) = -\, \dfrac{ \partial\Omega }{\partial\Phi } \,.
\end{equation}

On a ring the fields $\theta_{\rho}$ and $\theta_{\sigma}$ obey the following twisted boundary conditions \cite{Loss92, FK93}

\begin{eqnarray}
\dfrac{1}{\sqrt{\pi} }\, \theta_{\rho}(\tau + k_{2}\beta, x + k_{1}L) &=& \dfrac{1}{\sqrt{\pi} }\, \theta_{\rho}(\tau,x) \nonumber \\
&&  + k_2 n_{\rho} + k_{1} \left(2m_{\rho} + k_{M\rho}\right)\,, \nonumber \\
\label{eq06} \\
\dfrac{1}{\sqrt{\pi} }\, \theta_{\sigma}(\tau + k_{2}\beta, x + k_{1} L) &=& \dfrac{1}{\sqrt{\pi} }\,  \theta_{\sigma}(\tau, x) \nonumber \\
&& + k_{2} n_{\sigma} +  k_{1} \left(2m_{\sigma} + k_{M\sigma}\right) \,, \nonumber
\end{eqnarray}

\noindent
where $k_{1}$, $k_{2}$, $n_{\rho(\sigma)}$, and $m_{\rho(\sigma)}$ all are integers.
Moreover, both $n_{\rho}$ and $n_{\sigma}$ (and accordingly $m_{\rho}$ and $m_{\sigma}$) have the same parity.
The topological numbers are $k_{M{\rho}} = k_{M{\uparrow}} + k_{M{\downarrow}}$ and $k_{M{\sigma}} = k_{M{\uparrow}} - k_{M{\downarrow}}$.
The spin-resolved topological numbers $k_{M\uparrow/\downarrow}$ characterize the parity of the number of additional (over the ground state number) electrons in the ring.
In combination with previously introduced topological numbers $k_{j\uparrow/\downarrow}$ (dependent on the parity of the total number of electrons) one can relate them to the parity of the number $N_{0\uparrow/\downarrow}$ of electrons in the ground state in such a way that $k_{M\uparrow/\downarrow} = k_{j\uparrow/\downarrow}$ if $N_{0\uparrow/\downarrow}$ is odd and $k_{M\uparrow/\downarrow} = (k_{j\uparrow/\downarrow} + 1) \mod{1}$ if $N_{0\uparrow/\downarrow}$ is even. \cite{Loss92,FK93}

Since the Lagrangian under consideration is quadratic in the fields $\theta_{\rho/\sigma}$, the extremal trajectories obeying the boundary conditions (\ref{eq06}) and determining the flux-dependent part of the partition function $Z = A Z(\Phi)$ are linear functions of both $x$ and $\tau$:

\begin{equation}
\theta_{\rho/\sigma}^{ext}(\tau, x) = \sqrt{ \pi} \left\{ (2m_{\rho/\sigma} + k_{M{\rho/\sigma} } )\, \frac{x}{L} + n_{\rho/\sigma}\, \frac{\tau}{\beta} \right\} .
\label{eq07}
\end{equation}

The measure $D\theta_{\rho} D\theta_{\sigma}$ in Eq.~(\ref{eq05}) includes the summation over $n_{\rho/\sigma}$ and $m_{\rho/\sigma}$ that defines $Z(\Phi)$.
The integration over fluctuations of fields $\theta_{\rho/\sigma}$  defines a magnetic-flux independent constant $A$.

Since in the ground state the system under consideration is non-magnetic, $N_{0\uparrow} = N_{0\downarrow}$, then $N_{0}$ is even.
We calculate $Z(\Phi)$ for $N_{0} = 4n + 2$, where $n$ is an integer.
In this case we find, \cite{Moskalets00}

\begin{eqnarray}
Z(\Phi) &=& \sum\limits^{4}_{i=1} \theta_i(2\varphi_{AB},q)\, \theta_i(0,q) \, \theta_i(1,q_{c})\, \theta_i(2\varphi_{Z},q)\,, \nonumber \\
\label{eq08}
\end{eqnarray}

\noindent
where  $\varphi_{AB} = \Phi/\Phi_{0} \equiv B/\Delta B_{AB}$, $\varphi_{Z} = B/\Delta B_{Z}$, $q = e^{- \frac{2\pi^{2} T }{\Delta_{F} } }$, $q_{c} = e^{- \frac{2\pi^{2} T }{\Delta_{F} + 8 E_{c} } }$, and  $\theta_i(v,q)$ are the Jacobi theta functions, see, e.g., Ref.~\onlinecite{BE52}:

\begin{eqnarray}
\theta_{1}(v,q) &=& 2 \sqrt[4]{q } \sum\limits_{n=0 }^{\infty } (-1)^{n}\, q^{n(n+1) }\, \sin[ (2n + 1) \pi v] \,, \nonumber \\
\nonumber \\
\theta_{2}(v,q) &=& 2 \sqrt[4]{q } \sum\limits_{n=0 }^{\infty } q^{n(n+1) }\, \cos[ (2n + 1) \pi v] \,, \nonumber \\
\label{eq09} \\
\theta_{3}(v,q) &=& 1 + 2 \sum\limits_{n=1 }^{\infty } q^{n^{2} }\, \cos( 2 n \pi v)\,, \nonumber \\
\nonumber \\
\theta_{4}(v,q) &=& 1 + 2 \sum\limits_{n=1 }^{\infty } (-1)^{n}\, q^{n^{2} }\, \cos( 2 n \pi v)\,, \nonumber
\end{eqnarray}

\noindent
Note, that the partition function for $N_{0} =  4n$ can be deduced from that for $N_{0} = 4n+2$ by changing $\varphi_{AB} \to \varphi_{AB} + 1/2$.

The magnetic field $B$ enters  $\Omega(\Phi)$ in a twofold way.
First, it does through the parameter $\varphi_{AB}$
that causes conventional AB oscillations with period $\Delta B_{AB}$.
Second, it does through the parameter $\varphi_{Z}$ that also causes oscillations of the thermodynamic potential, hence persistent current oscillations, with period $\Delta B_{Z}$.
The ratio of corresponding periods can be represented as follows:

\begin{equation}
\label{eq10}
\dfrac{\Delta B_{Z} }{\Delta B_{AB} } = \dfrac{\Delta_{F} }{\Delta_{Z0} } \,,
\end{equation}

\noindent
with $\Delta_{Z0} = g\mu_{B} \Phi_{0}/S$ the Zeeman splitting at the magnetic field producing one magnetic flux quantum through the ring's opening.

Taking the gyromagnetic ratio $g =2$ and assuming the carrier's mass equal to a free electron mass, we find in the ballistic case under consideration, $\Delta B_{Z}/\Delta B_{AB} = N_{0}/4$. \cite{Moskalets99}.
The factor $1/4$ reflects the well known parity effect \cite{Leggett91}$^{,}$ \cite{Loss92} for spinful electrons.
For $N_{0} \gg 1$, the period of oscillations caused by the Zeeman splitting is much larger than the period of AB oscillations.
Therefore, the former effect will result in a large-scale modulation (a beating) of the AB oscillations.

\begin{figure}[t]
\centerline{\psfig{file=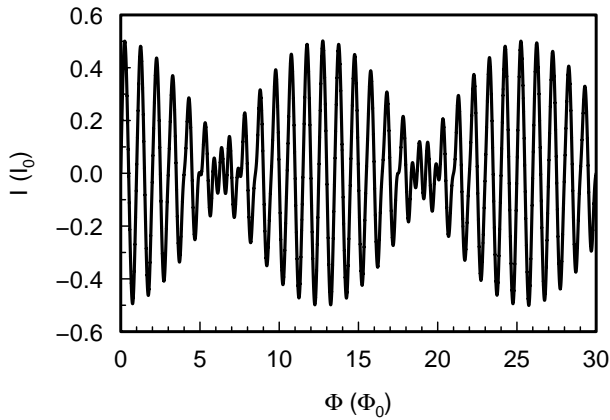,width=8cm}}
\caption{Persistent current $I$ in units of $I_{0} = ev_{F}/L $ as a function of the magnetic flux $\Phi = B S$. The parameters are: $k_{B} T = 0.2 \Delta_{F}$, $E_{c} = 10\Delta_{F}$,  $\Delta_{Z0} = 0.04 \Delta_{F}$.}
\label{fig2}
\end{figure}

In Fig.~\ref{fig2} we give the persistent current, Eq.~(\ref{pc}), calculated using the partition function, Eq.~(\ref{eq08}).
The suppression of the persistent current $I = I_{\uparrow} + I_{\downarrow}$ occurs for magnetic fields at which the number of spin excitations in the ring, $N_{\sigma} = N_{\uparrow} - N_{\downarrow}$ changes.
Strictly speaking at these fields $N_{\sigma}$ fluctuates. Since the total number $N_{e}$ of electrons is fixed, it fluctuates by $\pm 2$.
The numbers $N_{\uparrow}$ and $N_{\downarrow}$ fluctuate by $\pm 1$. Because of the parity effect, that results in fluctuations of an effective magnetic flux experienced by electrons by $\Phi_{0}/2$.
Correspondingly the currents $I_{\uparrow}$ and $I_{\downarrow}$ fluctuate such that the odd harmonics of their dependence on $\Phi$ vanish.
Therefore, the Zeeman effect results in halving of the period of AB oscillations at some particular fields.
With increasing temperature the higher harmonics decay faster and only the first one has noticeable magnitude. \cite{CGRS88, Loss92}
This is the reason why the period halving appears together with quenching of the current in Fig.~\ref{fig2}.

Note the mechanism of period halving we discuss here is different from the general one inherent to the system of electrons with spin discussed in Ref.~\onlinecite{LG91}.

So, quenching of the current in high magnetic fields (accordingly to the mechanism presented here) does not mean that the persistent current is destroyed.
And the subsequent revival of the current shows that the system remains phase-coherent.
The high temperature may act similarly: It suppresses a current leaving an electron system in the phase-coherent state that can be revealed with the help of persistent current fluctuations.

\section{Thermal fluctuations of persistent currents in ballistic rings}
\label{sec3}
Due to influence of the thermal bath the persistent current fluctuates.
The fluctuations, like the persistent current, exist only if the electron state is phase coherent.
Any dephasing processes destroy both the persistent current and its fluctuations.
However, investigating a temperature behavior of the persistent current only, it is difficult to say why it is destroyed, because of strengthening of decoherence processes with temperature or because of a mutual cancellation of contributions from thermally excited states.
Therefore, detecting the thermal fluctuations of the persistent current could shed more insight onto the quantum state of electrons in a ring at non-zero temperatures.

\subsection{Thermodynamic approach to persistent current fluctuations}

First we start from a two-level model which can be used to describe electrons in a ring at low temperatures. \cite{LB85, MButtiker85}
Then we present calculations with the total spectrum taken into account that is necessary at high temperatures.

\subsubsection{Two-level model}

Here we follow Ref.~\onlinecite{BJ05}.
Consider a temperature so low that only the first elementary excitation is important. In a canonical system we have a probability $p_{-}$ that the system is in the ground state and $p_{+}$ is the probability that the topmost electron of the ground state has been excited into the first available state. The energies of the two states are,

\begin{equation}
\label{eqt1}
E_{-}= \sum_{n=1}^{n = N} \epsilon_{n}\,,
\end{equation}

\noindent
and

\begin{equation}
\label{eqt2}
E_{+}= \epsilon_{N+1} + \sum_{n=1}^{n = N-1} \epsilon_{n} \,.
\end{equation}

\noindent
We have assumed that there are $N$ spinless electrons. In the Boltzmann case we have $p_{+}= (1/Z) \exp\left\{-E_{+}/(k_{B} T) \right\}$ and $p_{-}= (1/Z) \exp\left\{- E_{-}/(k_{B} T) \right\}$.
The normalization $Z$ is determined by the requirement that $p_{+} + p_{-}= 1$, and thus

\begin{equation}
\label{eqt3}
p_{+}= \frac{1}{e^{ \frac{\Delta E}{ k_{B} T } } + 1 }
\end{equation}

\noindent
and

\begin{equation}
\label{eqt4}
p_{-}= \frac{1}{e^{- \frac{ \Delta E }{ k_{B} T} } + 1 }
\end{equation}

\noindent
where $\Delta E = E_{+} - E_{-} = \epsilon_{N+1} - \epsilon_{N}$.

The currents in these two states are $I_{+} = \partial E_{+}/\partial \Phi$ and  $I_{-} = \partial E_{-}/\partial \Phi$. The average current in the two state approximation is

\begin{equation}
\label{eqt5}
\left\langle I \right\rangle = I_{+}p_{+} + I_{-}p_{-} \,.
\end{equation}

\noindent
Note that at low temperatures the excited state probability is exponentially small $p_{+}= {1}/{\exp\left\{ \Delta E/(k_{B} T) + 1\right\} } \approx \exp\left\{-  \Delta E/(k_{B} T) \right\}$. Therefore the departure from the ground state is exponential. Of course here we assume that the levels are non-degenerate. If they are degenerate then $p_{+} = p_{-} = 1/2$ which is for instance the case for a ballistic ring either in the center or at the boundary of the Brillouin zone depending on whether the particle number is odd or even.

The mean of the square of the current is

\begin{equation}
\label{eqt6}
\left\langle I^{2} \right\rangle = I^{2}_{+}p_{+} + I^{2}_{-}p_{-}
\end{equation}

\noindent
and thus for the mean square current fluctuations we obtain

\begin{equation}
\label{eqt7}
\left\langle \delta I^{2}\right\rangle =  (I_{+} - I_{-})^2 p_{+}p_{-} \,.
\end{equation}

Note the that this expression is just what we expect for thermal fluctuations, since $f =  p_{-}$,
and $p_{+} = 1- f$ but of course here the $p's$ are not Fermi distribution functions. But it holds for the $p's$ like for the Fermi functions that $f(1-f) = - k_{B}T df/dE$. We can make the following statement about the temperature dependence: if $k_{B}T < \Delta E$ the temperature dependence is exponential: that is the mean square current fluctuations are exponentially small. If we are at a point of degeneracy then $I_{+}= - I_{-} = I_{0}$ and $p_{+} = p_{-} = 1/2$ and the mean square current fluctuations are

\begin{equation}
\label{eqt8}
\left\langle \delta I^{2}\right\rangle = I_{0}^{2} \,.
\end{equation}

\noindent
Note that in the two level approximation this represents an upper bound for the current fluctuations. For the ballistic ring at low temperatures the mean square current fluctuations should be a strongly varying function of flux (and temperature). Namely at the degeneracy points the above results should apply with a maximal fluctuations whereas away from the degeneracy points for $k_{B}T < \Delta E$ the fluctuations remain exponentially suppressed, see a blue dash-dotted line in Fig.~\ref{fig3}.

In Ref.~\onlinecite{BJ05} this model was used to discuss the case if the many-body quantum mechanical ground state of system and environment are entangled. In this case the current fluctuations persist down to zero temperature and provide entanglement information.

To account for a high-temperature behavior we generalize straightforwardly a two-level model.

\subsubsection{Multi-level model}

Let us assume that we know the multi-electron spectrum $E_{k}(\Phi)$ for electrons in the ring threaded by the magnetic flux $\Phi$ and contacted with a thermal bath with temperature $T$.
Then we can calculate the partition function (see, e.g., Ref.~\onlinecite{LL5}),

\begin{eqnarray}
Z(\Phi) &=& \sum_{k}^{} e^{ - \frac{E_{k}(\Phi) }{ k_{B} T } }\,,
\label{eq12}
\end{eqnarray}

\noindent
and the persistent current, $I(\Phi) =  k_{B} T \partial \ln Z(\Phi)/\partial \Phi$  \cite{BY61}.
It is easy to see that the current $I(\Phi)$ can be represented as an average of currents supported by individual levels, $I_{k}(\Phi) = -\, \partial E_{k}(\Phi)/\partial \Phi$, found with the help of  the Gibbs distribution function, $ w_{k}(\Phi) = Z(\Phi)^{-1} \exp\left\{-\, E_{k}/(k_{B} T) \right\} $ :
\begin{eqnarray}
I(\Phi) \equiv \left\langle I \right\rangle &=& \sum\limits_{k}^{} I_{k}(\Phi)\, w_{k}(\Phi)\,.
\label{eq13}
\end{eqnarray}

\noindent
By analogy we define the mean square current fluctuations as follows:

\begin{eqnarray}
\left\langle \delta I^{2} \right\rangle &=& \sum_{k}^{} \left[ I_{k}(\Phi) - I(\Phi) \right]^{2}\, w_{k}(\Phi)\,.
\label{eq15}
\end{eqnarray}

\noindent
Using Eqs.~(\ref{eq12}) - (\ref{eq15}), after a little algebra, we can connect the fluctuations and the persistent current,

\begin{eqnarray}
\left\langle \delta I^{2} \right\rangle = k_{B} T\, \left( \dfrac{\partial I(\Phi) }{\partial \Phi } + \gamma(\Phi) \right) , \nonumber \\
\label{eq16}  \\
\gamma(\Phi) = \sum_{k}^{} \dfrac{\partial^{2} E_{k}(\Phi) }{\partial \Phi^2 }  \, w_{k}(\Phi)\,. \nonumber
\end{eqnarray}

\noindent
Since we did not use the explicit expression for the multi-particle spectrum  $E_{k}(\Phi)$, the equations given above are valid for rings with either fixed number of electrons (canonical case) or fixed chemical potential (grand canonical case). \cite{Moskalets01}
Also they are valid for rings with disorder and with interactions.

At low enough temperatures only two lowest levels, say, $ k = 0, 1$, matter and we recover a two-level model with $w_{0} = p_{-}$ and $w_{1} = p_{+}$.

\subsubsection{Fluctuations for non-interacting electrons with fixed chemical potential}

In the particular case of a ring exchanging with a bath both energy and particles, the occupation of single-particle energy levels $\epsilon_{n}(\Phi)$ are given by the Fermi distribution function $f_{0}(\epsilon_{n})$ with a bath temperature $T$ and chemical potential $\mu$.
In this case averaging over the multi-particle spectrum $E_{k}(\Phi)$ with the Gibbs distribution function $w_{k}(\Phi)$ is identical (for non-interacting particles) to averaging over the single-particle spectrum $\epsilon_{n}(\Phi)$ with the Fermi distribution function $f_{0}(\epsilon_{n})$.
Therefore, instead of Eqs.~(\ref{eq13}) and (\ref{eq15}) we can write,

\begin{eqnarray}
I(\Phi) \equiv \left\langle I \right\rangle &=& \sum\limits_{n}^{} i_{n}(\Phi)\, f_{0}(\epsilon_{n}) \,, \label{eq20} \\
\nonumber \\
\left\langle \delta I^{2} \right\rangle &=& \sum\limits_{n}^{} \left[ i_{n}(\Phi) - I(\Phi) \right]^{2} f_{0}(\epsilon_{n})\,, \label{eq21}
\end{eqnarray}

\noindent
where $i_{n}(\Phi) = - \partial \epsilon_{n}(\Phi)/\partial \Phi$ is a single-electron current.
Then using Eq.~(\ref{eq16}) with $\gamma= \sum_{n} \partial^2 \epsilon_{n}/\partial \Phi^2\, f_{0}(\epsilon_{n})$ we find:

\begin{eqnarray}
\left\langle \delta I^{2} \right\rangle = \sum\limits_{n}^{} i_{n}^{2}(\Phi) \, \delta N_{n}^{2}(\Phi) \,, \label{eq21_1}
\end{eqnarray}

\noindent
where $\delta N_{n}^2(\Phi) = f_{0}(\epsilon_{n}) [1 - f_{0}(\epsilon_{n}) ]$ is the mean square fluctuations of the occupation number of a level with energy $\epsilon_n(\Phi)$.

In a ring with many electrons, $\mu \gg \Delta_{F}$, we can simplify Eq.~(\ref{eq21_1}) noting that only the levels close to the Fermi energy contribute to fluctuations (we assume $\mu \gg k_{B} T$).
For these levels the absolute value of a current is roughly the same.
Then we can write,

\begin{eqnarray}
\left\langle \delta I^{2} \right\rangle = i_{F}^{2}(\Phi) \, \left\langle \delta N^2 \right\rangle \,, \label{eq21_2}
\end{eqnarray}

\noindent
where $i_{F}(\Phi)$ is a current for an electron with Fermi energy,
$\left\langle \delta N^2 \right\rangle$ is the mean square fluctuations of the electron number in the ring:

\begin{eqnarray}
\left\langle \delta N^2 \right\rangle \equiv \sum\limits_{n}^{} \delta N_{n}^{2}(\Phi)  = \sum\limits_{n}^{} f_{0}(\epsilon_{n}) [1 - f_{0}(\epsilon_{n}) ]\,.
\label{eq21_3}
\end{eqnarray}

\noindent
Therefore, for rings with fixed chemical potential, either clean or with disorder, the thermal current fluctuations are due to fluctuations of the number of particles.
In contrast, for rings with fixed number of particles, for which we can use Eqs.~(\ref{eq15}), (\ref{eq16}), the current fluctuations are due to transitions of an entire electronic system between levels supporting different currents.

Below we illustrate this general consideration with some simple examples.
We start with a ballistic ring model.

\subsection{Ballistic ring with fixed number of electrons}

To clarify the effect of temperature as much as possible,
we analyze the simplest model, which includes $N_{0}$ spinless non-interacting ballistic electrons confined in a $1D$ ring.
This system is coupled to a thermal reservoir with temperature $T$, while the particle exchange is forbidden.
To describe this model we use the Lagrangian $L_{LL}$, Eq.~(\ref{eq01}), with $\theta_{\rho} = \theta_{\sigma} \equiv \theta$, and the Lagrangian $L_{AB}$, Eq.~(\ref{eq02}), without the spin current, $j_{\sigma} = 0$ and with $k_{j\rho} = 2k_{j}$.
The topological number is $k_{j} = 0$ if $N_{0}$ is odd and is $k_{j} = 1$ if $N_{0}$ is even.
The twisted boundary condition reads:

\begin{equation}
\dfrac{1}{\sqrt{\pi} }\, \theta(\tau + k_{2}\beta, x + k_{1}L) = \dfrac{1}{\sqrt{\pi} }\, \theta(\tau,x)  + k_2 n_{\theta} \,, \label{eq11}
\end{equation}

\noindent
where $k_{1}$, $k_{2}$, and $n_{\theta}$ all are integers.
The extremal trajectories obeying this boundary condition, $\theta^{ext}(\tau,x) = \sqrt{\pi}\, n_{\theta} \tau/\beta$, define the magnetic-flux dependent factor of the partition function $Z(\Phi)$.
For odd $N_{0}$ it is $Z(\Phi) = \theta_{3}\left( \varphi_{AB},\, e^{-\, \frac{\pi^2 T }{\Delta_{F} } } \right)$. \cite{Loss92}
For even $N_{0}$ we should replace $\varphi_{AB} \to \varphi_{AB} + 1/2$.
Using the Poisson summation formula one can rewrite $Z(\Phi)$ as follows, \cite{Moskalets01}

\begin{eqnarray}
Z(\Phi) &=& \sum_{k=-\infty}^{\infty} e^{ - \frac{E_{k}(\Phi) }{ k_{B} T } }\,, \nonumber \\
\label{eq12_1} \\
E_k(\Phi) &=& \Delta_F\left( k + \frac{\Phi}{\Phi_0} +\, \frac{N_0 - 1}{2}\!\!\!\!\mod{1} \right)^2 . \nonumber
\end{eqnarray}

\noindent
Here $E_{k}(\Phi)$ is the spectrum of the system of $N_{0}$ (non-interacting and spinless) electrons in the ring.
It is easy to see, that Eq.~(\ref{eq12_1}) is not changed under the magnetic flux reversal, $\Phi \to -\,\Phi$. Therefore, the partition function, hence the free energy, is an even function of $\Phi$.

Calculating the relevant free energy, $F(\Phi) = - k_{B} T \ln Z(\Phi)$, and the  persistent current $I(\Phi) = - \partial F(\Phi)/\partial \Phi$, we can see that Eq.~(\ref{eq13}) holds.
After simple transformations one can get: \cite{Loss92}

\begin{eqnarray}
I(\Phi) &=& \dfrac{2\pi k_{B} T }{\Phi_{0} }\, \sum\limits_{m = 1}^{\infty} (-1)^{m N_{0} } \dfrac{\sin \left(2\pi m \frac{\Phi }{\Phi_{0} } \right) }{\sinh\left(m\, \frac{k_{B} T }{\Delta_{F}/\pi^{2} } \right) }\,. \label{eq13_1}
\end{eqnarray}

\noindent
Changing $N_{0}$ by $1$ is equivalent to changing $\Phi$ by $\Phi_{0}$.
This is a manifestation of the parity effect mentioned above.

At zero temperature the system is in its ground state which has energy $E_{0}$ at $\Phi = 0$.
In the ground state we have $w_{0} = 1$ and $w_{k} = 0$ for $k \ne 0$.
At non-zero temperature the system is excited to higher states, $w_{k} \ne 0$.
Therefore, at non-zero temperature many states do contribute to the current.
This leads to persistent currents fluctuations that can be understood as follows.
The probability $w_{k}$ characterizes how long on average the system stays in the state with energy $E_{k}$.
While in this state there is a current $I_{k}$ flowing in the ring.
After (on average) the time period $\tau_{k} = C w_{k}^{-1}$ the system jumps into another state, say, with energy $E_{k^\prime}$ and the circulating current changes to $I_{k^\prime}$.
Due to these changes the current does fluctuate.
These are classical (or quasistationary in terminology of Ref.~\onlinecite{LL5}) fluctuations.
Since the persistent current is quantum, we conclude that the fluctuations under discussion are classical (quasistationary) fluctuations of the quantum quantity.

Note that the proportionality factor $C$ defining the scale of time depends crucially on the strength of coupling between the ring and the thermal reservoir.
With increasing coupling, when $\tau_{k}$ becomes comparable with the time of a single turn around the ring, this classical approach should fail and the fluctuations should be treated quantum-mechanically.

For a ballistic ring with fixed number of electrons we use the spectrum $E_{k}(\Phi)$ given in Eq.~(\ref{eq12_1}) and find:

\begin{equation}
\label{eq17}
\left\langle \delta I^{2} \right\rangle = k_{B} T\, \left( \dfrac{\partial I(\Phi) }{\partial \Phi } + \dfrac{2\Delta_{F} }{\Phi_{0}^{2} } \right) .
\end{equation}

\begin{figure}[t]
\centerline{\psfig{file=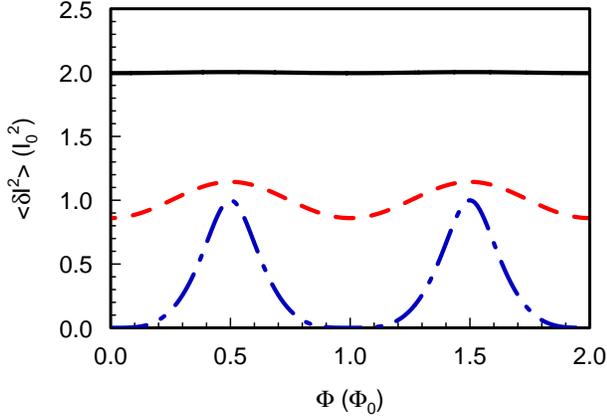,width=8cm}}
\caption{(Color online) Mean square fluctuations of the persistent current in the ballistic ring $\left\langle \delta I^{2} \right\rangle$ in units of $I_{0}^{2} =( ev_{F}/L)^{2} $ as a function of the magnetic flux $\Phi$. The temperature is $k_{B} T = \Delta_{F}$ (black solid line), $0.5\Delta_{F}$ (red dashed line),  $0.1 \Delta_{F}$ (blue dash-dotted line). The number of particle in the ring is odd.}
\label{fig3}
\end{figure}

\indent
The mean square current fluctuations is given in Fig.~\ref{fig3} for different temperatures.
At low temperatures, a blue dash-dotted line in Fig.~\ref{fig3}, the fluctuations depend on a magnetic flux.
Their maximal value agrees with Eq.~(\ref{eqt8}).
While with increasing temperature fluctuations become insensitive to a magnetic flux.
In particular at $T \gg \Delta_{F}/\pi^2$, when the persistent current vanishes, $I(\Phi) \approx 0$, the equation (\ref{eq17}) leads to a linear in temperature mean square fluctuations:

\begin{equation}
\left\langle \delta I^{2} \right\rangle = 2 I_{0}^2\, \dfrac{k_{B} T}{  \Delta_{F} } \,.
\label{eq17_1}
\end{equation}

\noindent
Interestingly, this result is also valid for a ballistic ring in the grand canonical case. \cite{Moskalets01}

The persistent current fluctuations in ballistic rings, Fig.~\ref{fig3}, show several rather counter-intuitive properties: (i) insensitivity to a magnetic flux at high temperatures and (ii) enhancing of the magnitude of mean square fluctuations at half of the magnetic flux quantum where the current vanishes.
These properties are a consequence of a direct level crossing inherent to a ballistic model.
Any small disorder would open gap at level crossings leading to vanishing of the fluctuations at both $\Phi = 0$ and  $\Phi = \Phi_{0}/2$.
To illustrate it we consider the next model.

\subsection{Single impurity in the ballistic ring with fixed chemical potential}

Let a  single  point  impurity  with  a potential $U(x) = g\delta(x)$ is embedded into a one-dimensional ballistic ring of a length $L$ with spinless non-interacting electrons coupled to a bath with temperature $T$ and chemical potential $\mu$.
The  eigenvalue  equation for an electron wave vector $k$ is (see, e.g., Ref.~\onlinecite{CGRS88}):

\begin{equation}
\cos\left(2\pi\, \dfrac{\Phi}{\Phi_0}\right) = \Re\left(\frac{\exp(- i kL)}{t} \right),
\label{eq18}
\end{equation}

\noindent
where $t = i\hbar v/( i\hbar v - g)$ (with $v = \hbar k /m_{e}$ a velocity) is a transmission coefficient through the potential $U(x)$.
For a strong potential, $|t| \ll 1$ we find a single-electron spectrum:

\begin{equation}
\epsilon_n (\Phi) = \frac{\pi^2\hbar^{2} n^{2} }{2m_{e}L^2}\left\{1 + (-1)^{n}\, \dfrac{2|t| }{\pi n } \cos\left( 2\pi\, \dfrac{\Phi}{\Phi_0} \right ) \right\} + {\cal O}(|t|^2) \,.
\label{eq19}
\end{equation}

\noindent
Here $n=1,2,\ldots$ is an integer, ${\cal O}(|t|^2)$ denotes small terms of order $|t|^2$  and higher.

Using Eq.~(\ref{eq19}) in Eq.~(\ref{eq21_1}) we calculate the mean square current  fluctuations in the ring with impurity (for $\mu \gg \Delta_{F},  k_{B}T$):
\begin{eqnarray}
\left\langle \delta I^2 \right\rangle &=& I_0^2 \left|t_{F}\right|^2 \sin^2\left(2\pi \dfrac{\Phi}{\Phi_0} \right) \left\langle \delta N^2 \right\rangle , \nonumber \\
\label{eq22} \\
\left\langle \delta N^2 \right\rangle &=& \frac{2T}{\Delta_F}\left( 1 + \frac{8\pi^2 k_{B} T}{\Delta_{F} }\sum_{q=1}^\infty\frac{q\cos\left( 2 q k_{F} L \right) }{ \sinh\left( q \frac{4\pi^2 k_{B}T }{\Delta_{F} } \right)} \right) , \nonumber
\end{eqnarray}

\noindent
where the lower index $F$ denotes quantities calculated at the Fermi energy, $\left\langle \delta N^2 \right\rangle$ is a mean square fluctuations of the number of electrons in the ring.
We stress here $\Delta_{F}$ is a level spacing near the Fermi energy in the similar but ballistic ring.
The level spacing in the ring with strong potential, $|t_{F}| \ll 1$, is two times smaller.

At high temperatures, $k_{B} T \gg \Delta_{F}/(4\pi^2)$, we have $\left\langle \delta N^2 \right\rangle \approx 2T/\Delta_F$ and the fluctuations grow linearly with temperature.
However in contrast to the ballistic case, Eq.~(\ref{eq17_1}), now, see Eq.~(\ref{eq22}), the fluctuations vanish at $\Phi=0$, $\Phi_{0}/2$ simultaneously with the vanishing of the persistent current.

To illustrate a crossover to the ballistic case we solve Eq.~(\ref{eq18}) at arbitrary transmission amplitude $t$ and get a spectrum close to the Fermi energy:

\begin{equation}
\epsilon_{n}^{(\pm)} (\Phi) \approx \frac{\Delta_{F} }{2n_{F} } \left\{n - \dfrac{\theta_{F} }{2\pi } \pm \dfrac{\arccos \left[ |t_{F}| \cos \left( 2\pi \dfrac{\Phi }{\Phi_{0} } \right) \right] }{2\pi }    \right\}^{2} \,,
\label{eq22_1}
\end{equation}

\noindent
where $n$ is chosen to be positive, $\theta_{F}$ the phase of the transmission amplitude: $t_{F} = |t_{F}|\, e^{i\theta_{F} }$, $n_{F}$ the serial number of the level closest to the Fermi level.

\begin{figure}[b]
\centerline{\psfig{file=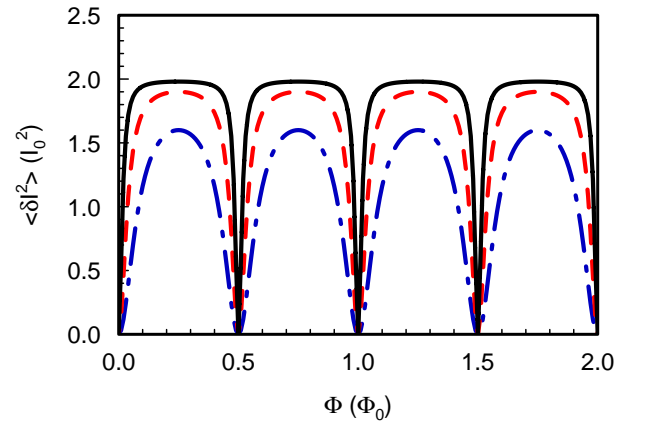,width=8cm}}
\caption{(Color online) Mean square fluctuations of the persistent current in the ring with impurity  $\left\langle \delta I^{2} \right\rangle$, Eq.~(\ref{eq23}), at a high temperature $k_BT = \Delta_{F}$ as a function of the magnetic flux $\Phi$ at different strengths of the reflecting potential. The reflection coefficient is $R = 0.01$ (black solid line), $0.05$ (red dashed line),  $0.2$ (blue dash-dotted line).}
\label{fig4}
\end{figure}

Then, using Eq.~(\ref{eq21_2}), we calculate the persistent current fluctuations at high temperatures, see Fig.~\ref{fig4}:

\begin{equation}
\left\langle\delta I^2 \right\rangle = 2I_0^2\, \frac{k_{B} T }{\Delta_F }\, \frac{ |t_F|^2\sin^2\left(2\pi \dfrac{\Phi}{\Phi_0} \right) }{ 1 - |t_F|^2\cos^2\left (2\pi \dfrac{\Phi}{\Phi_0} \right) } \,.
\label{eq23}
\end{equation}

\noindent
This equation reproduces both the ballistic case at $|t_{F}| = 1$, Eq.~(\ref{eq17_1}), and the case with a single strong impurity at $|t_{F}| \ll 1$, Eq.~(\ref{eq22}) at high temperatures.

So, from Fig.~\ref{fig4} one can see that the presence of even a weakly reflecting potential ($|t_F|\ll 1$) removes the counter-intuitive features characteristic for the persistent current fluctuations in purely ballistic rings.

\section{Conclusion}
\label{sec4}
Using a simple model, a one-dimensional ballistic ring with non-interacting electrons, we have shown several generic fluctuation effects.
We have considered effects of a high magnetic field and a high temperature on persistent currents.

With increasing magnetic field the Zeeman splitting leads to crossing of levels corresponding to electrons with opposite spins.
For an equidistant spectrum such crossing occurs periodically in magnetic field with period dictated by the level spacing for the ballistic ring.
At these particular magnetic fields the number of spin excitations in the ring fluctuates.
As a result the first harmonics of the persistent current becomes suppressed, hence the period of a current as a function of the Aharonov-Bohm flux through the ring's opening is halved.
At finite temperatures the magnitude of the second harmonics is generally smaller than the magnitude of the first one.
Therefore,  the period halving is accompanied by current quenching, see Fig.~\ref{fig2}.

With increasing temperature more and more excited energy levels in the ring are involved, hence the phase space accessible for an electron system is increased.
This results in a finite time spent by the system at some particular energy level, i.e., the position of a system in phase space fluctuates.
These fluctuations, first, affect the magnitude of the persistent current and, second, lead to fluctuations of the persistent current.
At high temperatures the magnitude of a current is exponentially suppressed.
However, the mean square current fluctuations in the presence of the Aharonov-Bohm magnetic flux grows linearly with temperature.
The existence of persistent current temporal fluctuations indicates that the system remains phase-coherent.

\begin{acknowledgments}

I am grateful to Markus B\"{u}ttiker for encouraging me to write this paper, for numerous fruitful discussions, and for careful reading of a manuscript.

\end{acknowledgments}

\end{document}